\documentclass[12pt]{article}
\let\d=\delta

\let\S=\Sigma

\newcommand{\nn}{\nonumber}
\newcommand{\un}{\underline}
\newcommand{\be}{\begin{equation}}
\newcommand{\ee}{\end{equation}}
\newcommand{\bea}{\begin{eqnarray}}
\newcommand{\eea}{\end{eqnarray}}

\renewcommand{\paragraph}[1]{
\vspace{.8mm}\par\noindent {\sl #1}\\
\vspace{0.2mm} }

\newcommand{\ft}[2]{{\textstyle\frac{#1}{#2}}}

\newcommand{\ba}{\left(\begin{array}}
\newcommand{\ea}{\end{array}\right)}

\begin{document}
\begin{flushright}
SU-ITP-98/63\\
KUL-TF-98/57\\
\hfill{hep-th/9812114}\\
\today\\
\end{flushright}

\begin{center}
{\large {\bf   Symmetries of
the Boundary of $AdS_{5}\times S^{5}$ \\

\

and Harmonic Superspace}}
\vskip 1 cm

{\bf  Piet Claus$^{+a}$, ~Renata Kallosh$^{*b}$ and J. Rahmfeld
$^{*c}$}\\
\vskip.5cm
{\small
$^+$
Instituut voor theoretische fysica, \\
Katholieke Universiteit Leuven, B-3001 Leuven, Belgium\\
\
$^*$ Physics Department, \\
Stanford University, Stanford, CA 94305-4060, USA\\
}
\end{center}

\vskip 1 cm We study the boundary limit of the bulk isometries of
$AdS\times S$. The superconformal symmetry is realized on the
coordinates of the $AdS$ boundary, the fermionic superspace
coordinates, and the harmonics on the sphere. We show how these
may be related to the coordinates of an off-shell harmonic
superspace of SCFT living on the boundary.
In the special case of $d=4$, ${\cal N}=2$ super Yang-Mills theory, a truncation of
the ${\cal N}=4$ SYM dual to Type IIB string theory compactified on
$AdS_5\times S^5$, we identify the bosonic space $SU(2)/U(1)$ of
the ${\cal N}=2$ harmonic superspace (known before as auxiliary) with the
$S^2$ submanifold of the $S^5$.

\vskip 0.2 cm
{\vfill\leftline{}\vfill
\footnoterule
\noindent
{\footnotesize
$\phantom{a}^a$ e-mail: Piet.Claus@fys.kuleuven.ac.be. }  \vskip  -5pt
\noindent
{\footnotesize
$\phantom{b}^b$ e-mail: kallosh@physics.stanford.edu. }  \vskip  -5pt

\noindent
{\footnotesize
$\phantom{c}^c$ e-mail: rahmfeld@leland.stanford.edu. }  \vskip  -5pt

\pagebreak
\setcounter{page}{1}

There is a growing interest in the $AdS$/CFT correspondence
conjecture \cite{Malda}. One of the basic arguments in favor of this
conjecture concerning ${\cal N}=4$ Yang-Mills theory is the fact that the
superisometry of the $AdS$ superspace forms the same ${SU(2,2|4)}$
supergroup as the symmetries of the ${\cal N}=4$ Yang-Mills field theory, which
supposedly lives on the boundary of the $AdS$ space.  The superspaces with
the bosonic $AdS_{p+2}\times S^{d-p-2}$ geometry and the proper form field
were recently constructed in \cite{NearHorizon}, and their
superisometries were found in closed form in \cite{CK}. The
purpose of this note is to motivate a connection between the limit to the
boundary ($\rho\rightarrow \infty$) of the bulk $AdS_{5}\times S^{5}$
superspace with the harmonic superspace \cite{GIKOS} of the off-shell
4-dimensional ${\cal N}=2$ Yang-Mills theory.

The superisometries found in \cite{CK} for the $AdS_{5}\times S^{5}$
case are {governed} by the supercoset structure
${G/H}=SU(2,2|4)/(SO(1,4)\times SO(5))$ \cite{Tseytlin, NearHorizon}.
They are of the form
\begin{eqnarray}
\delta \theta|_{\rm bulk} &=& \delta \theta (x, \rho, u, \theta\,|\,
\epsilon,  \eta,  a,
\lambda_M,  \lambda_D,  \Lambda_K,  \Lambda_R )\,, \\
\delta x |_{\rm bulk}&=& \delta x \left( x, \rho, u, \theta\,|\,
\epsilon,  \eta,  a, \lambda_M,  \lambda_D,  \Lambda_K,  \Lambda_R
\right)\,, \\
\delta \rho|_{\rm bulk} &=& \delta \rho  (x, \rho, u, \theta\,|\,
\epsilon, \eta,  a,  \lambda_M,  \lambda_D,  \Lambda_K,  \Lambda_R )\,,\\
\delta u|_{\rm bulk} &=& \delta u  (u, \theta  \, |\, \epsilon,
\eta, a, \lambda_M,  \lambda_D,  \Lambda_K,  \Lambda_R )\,.
\end{eqnarray}
Here, $x$ are the coordinates of the 4-dimensional space parallel to the
boundary of the $AdS_5$ space, and $\rho$ is the $AdS_5$ coordinate orthogonal
to the boundary, which is located at $\rho=\infty$. The 32
fermionic coordinates $\theta$ are the fermionic coordinates of type
IIB string theory/supergravity, and finally $u$ are harmonics of the five-sphere,
which may be given as functions of the angles parametrizing
the sphere, i.e.~$u(\phi)$.
The superisometries are rigid transformations, i.e. they are well defined
functions of the superspace coordinates and {\it global} parameters. These are
naturally the parameters of the superconformal transformations, namely
supersymmetry $\epsilon$, special supersymmetry
$\eta$, translations $a$, Lorentz transformations $\lambda_M$, dilatations
$\lambda_D$, special conformal transformations $\Lambda_K$ and
$R$-symmetry $\Lambda_R$.

Under these transformations (which have to be supplemented by the
compensating transformations of the frame from the stability group
$(SO(4,1)\times SO(5))$ the geometry of the $AdS_{5}\times S^{5}$
superspace is invariant.
The expressions for the superisometries $\d Z^M$ (where $Z$ denotes all
superspace coordinates) are rather complicated, and
they are explicitly constructed  in \cite{CK}. In short, they are given in terms
of  components of a covariantly constant Super-Killing field $\Sigma(Z)$ and
the inverse of the supervielbein $E_{\bar M}{}^M$ via\footnote{For details and
notations see \cite{CK}. Note that we changed a sign relative to (3.1) of
\cite{CK} for readability.}
\be
\d Z^M=\Sigma^{\bar M} E_{\bar M}{}^M\,.
\ee
At $\theta=0$, the supervielbein reduces to the standard vielbein, the gravitino,
and a fermion-fermion sector.
The Killing field $\Sigma$ depends linearly on the Killing
vectors and Killing spinors $\Sigma_0(x, \rho, u(\phi))$ of the background
geometry (and the corresponding
compensating transformations). Naturally, both $E$ and $\Sigma$ are also functions
of the $SU(2,2|4)$ structure constants and $\theta$.

At first glance the expressions for the isometries do not have a well
defined limit at $\rho\rightarrow \infty$.  For example, the metric, and hence the vielbein,
has terms with positive and negative powers of $\rho$:
\begin{equation}
g_{mn}   = \rho^2 \delta_{mn}  \qquad  g_{\rho \rho} =\left( { R\over
\rho}\right)^2 \,,
\end{equation}
where $R$ is the radius of curvature of the $AdS_5$ space as well as of
the sphere $S^5$.
Also, the translations $\Sigma_0^{m}$ parallel to the boundary  and
the compensating Lorentz transformations $\Sigma_0^{m \rho}$ which rotate in
the bulk-boundary plane $(x^m,\rho)$ have a non-trivial $\rho$ dependence.
They are of the form
\begin{equation}
\Sigma_0^{m} =  \rho \xi^m(x) + \left({R^2 \over \rho}\right)
\Lambda_K^m\,
\label{m}
\end{equation}
\begin{equation}
\Sigma_0^{m \rho} = \left({ \rho\over R}\right)\xi^m(x)
- \left({ R\over \rho}\right)\Lambda_K^m\,.
\label{compadS}
\end{equation}
In addition, we see the dependence on $\rho$ in the Killing spinors of the
$AdS_{5}\times S^{5}$ space, which depend on the
harmonics\footnote{We use a two-component notation, which means that we work
with a representation of $\gamma$-matrices in $5$ dimensions ($AdS$) for
which $\gamma_5$ is diagonal. The $AdS_5\times S^5$ Killing spinor consists of
four 4-dimensional Dirac spinors transforming under the $SU(4)$ with index
$i=1,2,3,4$ and the index $I=1,2,3,4$ labels the spinorial representation of
the stability group on the sphere $SO(5)$. More details on Killing spinors in
terms of the harmonics will be given in a future publication \cite{CKRZ}.}
of the five sphere $u_i{}^I(\phi)$:
\bea
\pmatrix{\S_{0 \ \alpha}^I \cr \bar \S_0^{\dot \alpha\, I}}
&=& \ba{c}\left({\rho\over
R}\right)^{1/2} \epsilon(x)_{\alpha}{}^i\\
\left({R \over \rho}\right)^{1/2} \bar \eta^{\dot\alpha\, i}\ea
u_i{}^{I}(\phi)\,,
\label{Killing}
\eea
where
\begin{equation}
\epsilon(x)_\alpha{}^i = \epsilon_{\alpha}{}^i + (x^m
\sigma_m)_{\alpha\dot\alpha} \bar\eta^{\dot\alpha\, i}\, ,
\end{equation}
and where $\epsilon$ and $\bar\eta$ are the parameters of the $Q$-supersymmetry
and $S$-supersymmetry resp.
We have found that this irregular dependence of the bulk isometries on
$\rho$ can be perfectly organized if we use two important technical tools.
\begin{itemize}

\item The superisometries in the bulk were found in the $AdS$ basis
of the superalgebra which treats {\it the bulk direction
on equal footing with the boundary directions} since there is a
manifest $SO(4,1)\times SO(5)$ symmetry, mixing bulk and boundary directions.
The limit to the boundary, however, should be performed in the
superconformal basis of the algebra which has only
manifest $SO(1,3)$ symmetry in the $AdS_5$ sector, reflecting the split of
the $AdS$ coordinates into $x$ and $\rho$.

\item We use the Killing gauge in the superspace \cite{NearHorizon},
since in this gauge the gravitino vanishes not only at $\theta=0$
for the bosonic background  but also to all orders in $\theta$ which
makes various calculations doable and easy.
In this gauge
we find the limit to the boundary $\rho\rightarrow \infty$ in which
the radial direction $\rho$ decouples.

\end{itemize}
The $SU(2,2|4)$ superalgebra with generators ${\bf T}_\Lambda$ and structure
constants $f_{\Lambda\Sigma}{}^\Delta$
\begin{equation}{}
[\, {\bf T}_\Lambda , {\bf T}_\Sigma\, \}  =
f_{\Lambda\Sigma}{}^\Delta {\bf T}_\Delta\,  \label{Galg}
\end{equation}
is defined up to linear redefinitions of the generators. The  most
useful basis for our purposes is the superconformal
basis, whose most
important property is that there exists a grading operator $D$, under which
{\it every generator  ${\bf T}_\Lambda$ has a particular conformal weight}:
\begin{equation}
[D, {\bf T}_\Lambda ] = W(T_\Lambda ) {\bf T}_\Lambda \,.
\end{equation}
In fact, the translation in the $\rho$-direction
becomes the dilatation operator $D$, translations along the boundary coordinates
$x$ combined with the Lorentz transformations
mixing the $\rho$ and $x$ directions give 4 translation and 4 special
conformal transformations along the boundary directions.
Translations $P_m$ have weight +1, special conformal transformations
$K_{m}$ have -1, dilatations $D$, Lorentz transformations $M_{mn}$
and R-symmetry transformations $U_i{}^j$ have weight 0. The supersymmetries
$Q$ have weight +1/2 and special supersymmetries $S$ have weight -1/2.
It turns out that the superconformal basis of the superalgebra
displays the $\rho$ dependence of isometries in a very clear way.
We have found that all isometries $\S$ at $\theta=0$ in the superconformal
basis are homogeneous in $\rho$, which is expected because the $\rho$
direction is associated with dilatations and every conformal operator has a
specific Weyl weight. They scale according to the conformal
weight of the corresponding operator:
\begin{equation}
\Sigma_0^\Lambda ( x^m, \rho, u(\phi) )\equiv
\rho^{W(\Lambda)}\un   \Sigma^{\Lambda} ( x^m,  u(\phi) ) =
\un \Sigma^{\un \Lambda}( x^m,  u(\phi) )
{\cal R}_{\un \Lambda}{}^\Lambda (\rho)\,.
\label{rho}\end{equation}
Now, the matrix ${\cal R}_{\un  \Lambda}{}^\Lambda(\rho)$
carries all dependence on the radial variable and the fields $\un \Sigma^{\un
\Lambda} (x^m, u(\phi))$ depend only on the bosonic coordinates of the
boundary, and the harmonics of the sphere.

The dependence of $\Sigma$ and $E$ on $\theta$ is such that use of
the  Killing spinor  gauge \cite{NearHorizon}
together with (\ref{rho}) allows
all non-trivial $\rho$ dependence in $\d Z$ to be absorbed into  new
$\rho$-dependent structure ``constants'' via the matrix ${\cal R}(\rho)$
defined in (\ref{rho}).
The crucial observation at this point is that the
structure constants of the algebra have the property that for any
component $f_{\Lambda \Sigma}^{\Delta}$
\begin{equation}
W(T_\Lambda) + W(T_\Sigma) = W(T_\Delta)
\label{rule}\end{equation}
holds.  This identity shows then that in the `underlined' basis the structure
``constants'' $f_{\un \Lambda \un \Sigma}^{\un \Delta} (\rho)$
are numerically nothing but the old constants $f_{\Lambda\Sigma}^{\Delta}$,
since
\bea
f_{\un \Lambda \un \Sigma}^{\un \Delta} (\rho)&\equiv &({\cal
R}^{-1})^{\un
\Delta}_{\Delta}(\rho)  {\cal R}^{ \Lambda}_{\un \Lambda} (\rho) \;
f_{\Lambda
\Sigma}^{\Delta} \; {\cal R}^\Sigma_ {\un \Sigma}(\rho)\nn \\
&=&
\rho^{W(T_\Lambda) + W(T_\Sigma) - W(T_\Delta)} \; \d^{ \Lambda}_{\un
\Lambda}
\d^{\un \Delta}_\Delta  \d_{\un \Sigma}^\Sigma \;
f_{\Lambda\Sigma}^{\Delta}
=\d^{ \Lambda}_{\un \Lambda}
\d^{\un \Delta}_\Delta  \d_{\un \Sigma}^\Sigma \;
f_{\Lambda\Sigma}^{\Delta}\,,
\eea
i.e.~the $\rho$-dependence cancels out. This gives us control over
the limit $\rho \rightarrow \infty$ in all expressions for superisometries. We
find that the bulk-boundary isometry correspondence has the following form:
\bea
\delta \theta|_{\rm bulk}& = & \delta \theta|_{\rm boundary}
(x,u,\theta)\,,\\
\delta x|_{\rm bulk} & = &\delta x|_{\rm boundary}(x,u,\theta)
+\frac{1}{\rho^2} f(x,u,\theta)\,,  \\
\delta u|_{\rm bulk}& = & \delta u|_{\rm boundary}(u,\theta)\,.
\eea
The isometry in the radial
direction blows up in the limit $\rho\rightarrow \infty$:
\begin{equation}
\delta \rho|_{\rm bulk}= \rho g(x,u,\theta) \, ,
\end{equation}
but decouples from the boundary isometries.

We do not specify here the functions of $(x,u,\theta)$ introduced above
as they are still rather complicated, but they can be read off from the
generic expressions for the isometries in the bulk,
given explicitly in \cite{CK}.

Thus we have extracted the  $\rho$-dependence of the isometries in the
superconformal basis and realized the algebra on coordinates of the
boundary space $(x,u,\theta)$.
In the bulk we have a spontaneously broken form
of the superconformal symmetry since $\rho$ occurs explicitly in the
transformations.
However at the boundary $\rho \rightarrow \infty$ where the $\rho$
dependence is absent, we have a realization of the superconformal symmetry
without spontaneous breaking.
Note that the boundary of the superspace naturally  includes the $4$ bosonic
coordinates of the boundary of the $AdS_5$ space, the fermionic
coordinates $\theta$, and {\it the harmonics of the $S^5$ sphere,
which are not decoupled at the boundary}!

One can now try to understand what the  role of these harmonics is from
the point of view of the SCFT in $4$ dimensions. Their structure and
transformation laws suggest a connection to the auxiliary coordinates of
the off-shell harmonic superspace of the $4$-dimensional Super Yang-Mills
theory. We will study the simplest example and {\it derive the ${\cal N}$=2 d=4
off-shell harmonic superspace} \cite{GIKOS} {\it of the super Yang-Mills theory
from the string theory/supergravity $AdS_5\times S^5$ superspace}.

To find the correspondence with the SYM theory we will require only  $Q$
supersymmetry to be realized manifestly.
The Lagrangians of the 4-dimensional ${\cal N}=2$ SYM theory have also a
superconformal symmetry but this does not have to be a manifest property
of the off-shell superfields.
To do the truncation consistently we do not require that the special
supersymmetry $S$, dilatation $D$, and the special conformal symmetry
$K$ are symmetries anymore. It means that we are fixing special
supersymmetry to keep the $\bar \lambda^{\dot \alpha i}$, which is the
component of $\theta$ along the special supersymmetry
direction\footnote{Here we use the {\it Killing Spinor gauge}
\cite{NearHorizon}, which means that we decompose
the 32 component fermionic coordinates $\theta$ in the same
way as the Killing spinor (\ref{Killing}), i.e.
$$
\theta_{\hat\alpha}{}^I = \ba{c}
\left({\rho\over R}\right)^{1/2} (\theta_{\alpha}^i + (x^m
\sigma_m)_{\alpha\dot\alpha} \bar \lambda^{\dot\alpha i})\\
\left({R \over \rho}\right)^{1/2} \bar \lambda^{\dot\alpha i}\ea
u_i{}^{I}(\phi)\,.
$$ },
at $\bar \lambda^{\dot\alpha i}=0$.
Thus we choose $\eta=0, \lambda_D = \Lambda_K=0$ simultaneously with
$\bar \lambda^{\dot \alpha i} = 0$.
The expressions for the isometries are greatly simplified, in particular,
the matrix ${\cal M}^2$, appearing in the isometries as given in \cite{CK},
drops out in the truncated theory.  We are left with 1/2 of the fermionic
coordinates, 4 bosonic coordinates $x^m$ and harmonics $u_i{}^I$ on $S^5$.
The fermionic coordinates related to $Q$-supersymmetry can be taken as
four complex 2-component left-handed $d=4$ spinors which we denote by
$\theta^i_{a}$, $i=1,2,3,4$ and $\alpha=1,2$.  We derive from the
superisometries in the bulk \cite{CK} that
after truncation the remaining symmetries at the boundary are:
\begin{equation}
\delta \theta^i_{\alpha} =   \epsilon^i_{ \alpha} +
\ft14 (\lambda_{mn}
\sigma ^{mn} \theta^i)_{ \alpha} +  \theta^j_{ \alpha}
\Lambda_j{}^i\,,
\label{dthi}
\end{equation}
\begin{equation}
\delta x^m = a^m + \lambda^{mn} x_n  + i (\epsilon^i  \sigma^m
\bar \theta_i - \theta^i  \sigma ^m
\bar \epsilon_i)\,.
\label{dx}
\end{equation}
The symmetries of the sphere are given in terms of the coset representative
(i.e. the harmonics $u(\phi)$) as follows
\begin{equation}
g u (\phi) = u (\phi') h(\phi,g)\,,
 \label{utransf}
\end{equation}
where,
$g$ is an element of $SO(6)$, $h(\phi)$ is an element of $SO(5)$, and
$\phi'$ are the transformed angles of the sphere. Explicitly, the
infinitesimal transformations take the form
\begin{equation}
-\Lambda_i{}^j u_j{}^K(\phi) = u_i{}^K(\phi+\delta\phi) - u_i{}^K (\phi)
- u_i{}^L(\phi) H_L{}^K(\phi,\Lambda)\,,
\end{equation}
where $\Lambda_i{}^j$ are the parameters of the transformation from
the $SO(6)$ group in spinorial representation
and $H_L{}^K(\phi)$ are the parameters of the
transformation from the stability group $SO(5)$.

The truncated system of coordinates at the boundary has a symmetry
realizing the truncated $SU(2,2|4)$ algebra without $S, D, K$, which is
the simply the extended ${\cal N}=4$ super Poincar\'{e} algebra in $d=4$. It
is realized on 4-dimensional coordinates $x^m$, on complex spinors
$\theta^i_{\alpha}$ and on harmonics $u_i{}^I(\phi)$ of the $S^5$
depending on 5 angles of the sphere.

Now, we would like to perform a further truncation of our boundary
superspace to reproduce the $d=4$ ${\cal N}=2$ harmonic superspace of GIKOS.
To do so we have to restrict ourselves to an $S^2$ subsphere of the full
$S^5$ and reduce the supersymmetries.
The symmetries of the reduced system then are:
\begin{equation}
\delta \theta^i_{\alpha} =   \epsilon^i_{ \alpha} +
\ft14(\lambda_{mn}
\sigma ^{mn} \theta^i)_{ \alpha}+  \theta^j_{ \alpha}
\Lambda_j{}^i\,,
\end{equation}
\begin{equation}
\delta x^m = a^m + \lambda^{mn} x_n  + i( \epsilon^i  \sigma ^m
\bar \theta_i - \theta^i  \sigma ^m
\bar \epsilon_i)\,,
\end{equation}
\begin{equation}
\delta u_i{}^K=  - \Lambda_i{}^j u_j{}^K +
u_i{}^L H_L{}^K\,.
\end{equation}
where  $\Lambda_i{}^j$ with $i,j=1,2$ are now the parameters of the
$SO(3)\sim SU(2)$ subgroup of $SO(6)$ and $H_I{}^J\equiv
\Lambda(\sigma_3)_I{}^J$ ($I, J=1,2 $) is the parameter of the
transformation of the stability subgroup $SO(2)\sim U(1)$. These are
{\it precisely the super Poicar\'{e} transformations of the harmonic
superspace in the central basis, as given in }\cite{GIKOS}.
This observation constitutes our main result.

The harmonics, introduced in \cite{GIKOS} as auxiliary $S^2$
coordinates\footnote{The new bosonic coordinates relevant for extended
supersymmetries were added to the superspace in the form of supertwistors
and/or coset coordinates in \cite{FerLuk}.
But the origin  of these additional coordinates which were used in
\cite{GIKOS} to construct an off-shell SYM theory was not clear so far to the
best of our understanding. Now we seem to shed some light also on the physical
origin of the {\it supertwistors} which may be related to the harmonic
superspace.}, are coset representatives of $SU(2)/U(1)$. A
common parametrization is
\begin{equation}
u(\phi_1, \phi_2) = \left |\matrix{
u_1^- & u_1^+\cr
 u^-_2& u_2^+ \cr
}\right | = \left |\matrix{
 \cos {\phi_1} & i e^{-i\phi_{2} }  \sin{\phi_1} \cr
i e^{i\phi_{2} }  \sin {\phi_1} &  \cos{\phi_1} \cr
}\right |\,.
\end{equation}
They also appear explicitly in the Killing spinors on the
$AdS\times S$ space \cite{CK}.
Their main purpose is to allow for a change of fermionic variables from
those transforming under $SU(2)$ to those transforming under the $U(1)$
\begin{equation}
\theta_{\alpha}^{\pm} \equiv  \theta_{\alpha}^{i}
u_i^{\pm}\,,
\end{equation}
where we took $I=(+, -)$. It follows that
\begin{eqnarray}
 \delta \theta^{\pm}_{ \alpha}=\epsilon^i_{ \alpha} u_i{}^{\pm}+
\ft14 (\lambda_{mn} \sigma^{mn}
\theta^{\pm})_{ \alpha} \pm  \theta^{\pm}_{ \alpha} \Lambda\, .
\label{dthpm}
\end{eqnarray}
The last term shows that $\theta^{\pm}$ transforms under the
$U(1)$ symmetry. This term came from the cancellation of the $SU(2)$
transformation from the $\theta^i$ by the shift of angular variables and the
remaining compensating $U(1)$ transformation is the one with the parameter
$\Lambda$.

The ABC of the analytic ${\cal N}=2$ superspace (which is
reminiscent of the chiral subspace of ${\cal N}=1$
susy) follows as in \cite{GIKOS}.  One can shift the real 4-dimensional
coordinates $x$ by the bilinear combination of spinors so that there is an
analytic subspace $x', \theta^+, \bar \theta^+, u$. Then one can
construct unconstrained superfields which depend in this basis only on
$\theta^{+}$. The complete ${\cal N}=2$ SYM theory with manifest off-shell
supersymmetry is then easily constructed. The SYM prepotential $V^{++}$
is an unconstrained analytic superfield of $U(1)$ charge $+2$, and the
hypermultiplets $q^+$ are given by analytic superfields of charge $+1$.

The basic difference between \cite{GIKOS} and present derivation lies in
the fact that we have the coordinates of $S^2$ as part of the real
boundary of the curved $AdS_5 \times S^5$ space.

Also, the somewhat tedious integration rules for the integration
over $u$, $\int du \cdot 1=1$ and $\int du\cdot  u^+_{(i}  u^-_{j)}=0$,
find a simple explanation in the $AdS_5\times S^5$ picture:
they all can be derived from
\begin{equation}
du = {1\over 4\pi} \sin \phi_1 d\phi_1 d\phi_2\, ,
\end{equation}
i.e.~from the integration over the physical sphere part of our boundary space.

This shows that the $S^2$  part of $S^5$ on which we compactify the
superstring/supergra-
vity theory plays a fundamental role in the existence of the
off-shell hypermultiplet action and supergraph Feynman rules of ${\cal N}=2$ SYM
theory developed in \cite{GIKOS}.  We find here the long-anticipated presence
of higher dimensions, however, they are not coming from the $d=10$ SYM
theory, but from string theory/supergravity on $AdS_5\times S^5$
through the $AdS/SCFT$ correspondence.
The harmonics of \cite{GIKOS}  find their place as the harmonics of the sphere
which is part of the boundary of the $AdS_5\times S^5$ space.

It would be very interesting to see whether a new (off-shell?) ${\cal N}=4$ harmonic
superspace can be constructed from  the untruncated boundary isometries
when the harmonics on the the full $S^5$ sphere are used.

\section*{Acknowledgements}
We are grateful to S.~Shenker for the suggestion to look at the
boundary limit of the  bulk isometries, to S.~Shenker and L. Susskind and
Y. Zunger for insightful discussions and to J.~Maldacena for asking us
about the role of the angles of the $S^5$ for the Yang-Mills theory in $d=4$
which lead us to identification of the boundary superspace with
the harmonic superspace. The work of R.~K. and J.~R. is supported by the NSF
grant PHY-9870115. The work of P.~C. is supported by the European
Commission TMR programme ERBFMRX-CT96-0045.




\begin{thebibliography}{10}
\bibitem{Malda}
J.~Maldacena,  {\sl The large n limit of superconformal field theories and
  supergravity}, {\tt hep-th/9711200} (1997);
E.~Witten,
 {\sl Anti-de sitter space and holography},
 {\tt hep-th/9802150} (1998);
S.~S. Gubser, I.~R. Klebanov and A.~M. Polyakov,  {\sl Gauge theory
correlators
from noncritical string theory}, Phys. Lett. B428 (1998) 105.
\bibitem{NearHorizon}
R.~Kallosh, J.~Rahmfeld and A.~Rajaraman,
\newblock {\sl Near horizon superspace}, J.High Energy Phys.
9809:002, 1998,
\newblock {\tt hep-th/9805217}
\bibitem{CK}
P.~Claus and R.~Kallosh,
\newblock {\sl Superisometries of the ads x s superspace},
\newblock {\tt hep-th/9812087} (1998).
\bibitem{GIKOS}
A.~Galperin, E.~Ivanov, S.~Kalitzin, V.~Ogievetsky and E.~Sokatchev,
\newblock {\sl Unconstrained N=2 matter, Yang-Mills and supergravity
theories in harmonic superspace}, CQG, {\bf 1}, 469 (1984).
\bibitem{Tseytlin}
R.~R. Metsaev and A.~A. Tseytlin,
\newblock {\sl Type iib superstring action in ads(5) x s(5) background},
\newblock Nucl. Phys. B533 (1998) 109.
\bibitem{CKRZ} P.~Claus, R.~Kallosh, J.~Rahmfeld, Y. Zunger, in
preparation.
\bibitem{FerLuk} A. Ferber,
 Nucl. Phys. B132 (1978) 55;  J. Lukierski,  in Proceedings of 1981
Karpacz
School (Wroclaw, 1982).
\end{thebibliography}

\end{document}